\documentstyle [12pt,aaspp4]{article}

\begin{document}

\title{Spectrophotometry of 237 stars in 7 Open Clusters}

\author {Lori Clampitt\\ and\\ David Burstein}

\affil {Department of Physics and Astronomy,\\ Arizona State University,\\
Tempe, AZ 85287-1504\\Email: clampitt@accesscom.com; 
burstein@samuri.la.asu.edu}

\vspace*{11cm}

\received{25 October 1996}
\accepted{May 13, 1997}

\begin{abstract}

Spectrophotometry is presented for 237 stars in 7 nearby open clusters:
Hyades, Pleiades, Alpha Persei, Praesepe, Coma Berenices, IC~4665, and M39.
The observations were taken by Lee McDonald and David Burstein using the
Wampler single-channel scanner on the Crossley 0.9m telescope at Lick
Observatory from July 1973 through December 1974.  Sixteen bandpasses spanning
the spectral range $\rm 3500 \AA$ to $\rm 7780 \AA$ were observed for each
star, with bandwidths 32$\rm \AA$, $\rm 48 \AA$ or $\rm 64 \AA$.  Data are
standardized to the Hayes--Latham system to mutual accuracy of 0.016 mag
per passband.  

The accuracy of the spectrophotometry is assessed in three ways on a 
star-by-star basis.  First, comparisons are made with previously published
spectrophotometry for 19 stars observed in common.  Second, (B--V) colors
and {\it uvby} colors are compared for 236 stars and 221 stars, respectively.
Finally, comparsions are made for 200 main sequence stars to the spectral
synthesis models of Kurucz, fixing log g = 4.0 and [Fe/H] = 0.0, and only
varying effective temperature.  The accuracy of tests using {\it uvby} colors
and the Kurucz models are shown to track each other closely, yielding
an accuracy estimate ($1 \sigma$) of 0.01 mag for the 13 colors formed from
bandpasses longward of the Balmer jump, and 0.02 mag for the 3 colors formed
from the three bandpasses below the Balmer jump.  In contrast, larger scatter
is found relative to the previously published spectrophotometry of 
Bohm-Vitense \& Johnson (16 stars in common) and Gunn \& Stryker (3 stars).

We also show that the scatter in the fits of the spectrophotometric colors
and the {\it uvby} filter colors is a reasonable way to identify the 
observations of which specific stars are accurate to $1 \sigma$, $2 \sigma$,
$\ldots$.  As such, the residuals from both the filter color fits and
the Kurucz model fits are tabulated for each star where it was possible to
make a comparison, so users of these data can choose stars according to the
accuracy of the data that is appropriate to their needs.

The very good agreement between the models and these data verifies the
accuracy of these data, and also verifies the usefulness of the Kurucz models
to define spectrophotometry for stars in this temperature range ($>$5000 K). 
These data define accurate spectrophotometry of bright, open cluster stars
that can be used as a secondary flux calibration for CCD-based 
spectrophotometric surveys.

\end{abstract}


\section{Introduction}
\label{intro}

The existing spectrophotometry available in computer-accessible form (The
Stellar Spectrophotometric Atlas, Gunn \& Stryker 1983; hereafter SSA; The
Catalog of Stellar Spectrophotometry, Adelman et al. 1989; hereafter CSS)
tends to concentrate either on stars for stellar synthesis work, or stars of
special interest (e.g., Ap stars; see Adelman et al.).  Spectrophotometry of
other stars also exists (e.g., Breger, 1976; Bohm-Vitense \& Johnson 1977
(hereafter BVJ); Ardeberg \& Virefors 1980), but these data are not yet in
computer-readable format.  Even then, only the study of BVJ explicitly tried
to sample stars in nearby open clusters.  A bit earlier than the BVJ
observations were taken, Lee McDonald and one of the authors (DB) decided to
obtain spectrophotometric data for a set of open cluster stars.  During
1973--4, 237 open cluster stars were observed using a scanning grating and a
photomultiplier detector; an instrument no longer in general use.  Any such
observations today are likely obtained with CCD-based instrumentation.

Given the different detector systematics of photomultipliers and CCDs, and
given the resurgence of interest in obtaining accurate fluxed data on stars
and galaxies with CCDs, these kind of data are still of interest to many
workers.  Of the 237 stars in our survey, only 25 stars are in common with
previous studies.  However, 236 stars have BV colors and 221 have
uvby colors to which we can quantitatively compare our results.  In addition,
spectrophotometry of such a large set of open cluster stars with a range of
age but otherwise similar metallicity is a good test for the Kurucz (1992,
1993) stellar synthesis models.

The spectrophotometry for 237 stars, using 16 bandpasses from 3600--7800$\rm
\AA$ is presented in \S~\ref{sec2}.  The seven open clusters observed are
Hyades (Hya), Pleiades (Ple), Alpha Persei (Per), Praesepe (Pra), Coma
Berenices (Com), IC~4665 (IC4), and M39.  Comparisons to published
spectrophotometry are also given in \S~\ref{sec2}.  Comparisons to Johnson
B--V and Str\"{o}mgren {\it u--v}, {\it v--y} and {\it u--b} colors are made
in \S~\ref{sec3}. The comparison to the models of Kurucz is given in
\S~\ref{sec4}.  We show that these comparsions reasonably lead to a 
quantitative means of assessing the accuracy of spectrophotometry on a 
star--by--star basis in \S~\ref{sec5}. Our summary is given in \S~\ref{sec6} 
and individual stars of special interest are discussed in Appendix~A.

\section{Observations, Calibrations and Data}
\label{sec2}

\subsection{Observations}
\label{sec2.1}

The observations of these open cluster stars were taken by Lee McDonald and
David Burstein during the period July 1973--Dec 1974, with the Wampler (1966)
scanning spectrometer attached to the 0.9m Crossley telescope at Lick
Observatory.  Table~1 lists the clusters observed in order of increasing age.
Parameters listed include Galactic latitude (b) and longitude (l), nominal
distance and distances estimated here using a fiducial zero-age main sequence
line (d), the value of (E(B-V)) used here, and estimated metallicity ([Fe/H]).
Table~2 gives the observing log, also noting the secondary standard star used 
for each night (see discussion below).

A one--channel ITT FW--130 photomultiplier coldbox was used every night except
the first, when a two--channel coldbox was used instead. Three filters
were used to separate the first and second orders of the gratings, and an
observation was made of each star in each filter. The effective wavelengths
used for passbands per filter are: Filter 1 (3500\AA, 3571\AA, 3636\AA,
4036\AA, 4167\AA, 4255\AA), Filter 2 (4167\AA, 4255\AA, 4400\AA, 4565\AA,
4785\AA, 5000\AA, 5263\AA), and Filter 3 (5263\AA, 5840\AA,  6300\AA, 6710\AA,
7100\AA, 7780\AA). Each passband within a filter set was measured twice,
with the starting wavelength measured a total of 4 times.  Note that three
passbands (4167, 4255, and 5263\AA) are observed in two separate filters. 
Multi-observations of a given filter are averaged together.

The passbands chosen are among the standard passbands of Oke (1964) and Hayes
(1970) which give a fairly uniform distribution in $1/\lambda$. During 1973,
all bandwidths for the observed passbands were 32$\rm \AA$.  In 1974, the
bandwidth for filters 1 and 2 (as defined above) were 48$\rm \AA$, while the
bandwidth used for filter 3 was 64$\rm \AA$.  Object observations were
interweaved with sky observations for all stars.  Hayes (1970) standard stars
were also observed during most nights, sampling among all 12 Hayes standards
with emphasis on observing $\alpha$ Lyrae.  At least 10,000 counts were
accumulated at each passband for each star to obtain sufficient signal to
noise.

\subsection{Calibration}
\label{sec2.2}

The first step in the data reduction (done in 1974) was the interpolation of
the sky measurements to the time of observation of a given star, which was
never significant at $>$1\% in the final magnitudes. Next, the observations
for the Hayes standard stars and the cluster stars for each night were reduced
using standard Mt. Hamilton extinction coefficients provided by E. J. Wampler.

One star from each cluster was measured several times during a given night
(such stars are listed in Table~2 as the ``secondary standard'') to internally
standardize each cluster onto the same system.  These stars were also used to
determine any trend of atmospheric extinction different from the assumed
coefficients during the night.  Only during Night 3 was a deviation detected,
affecting only the data for two stars (Per 285A, Per 383).  In addition, we
later discovered that two of the cluster secondary standard stars (Praesepe
300 and Coma 60) are Am stars, with one (Coma 60) suspected of being variable.
 Fortunately, we can explicitly test if use of these stars has affected the
calibration of these two clusters (\S~\ref{sec3}). To complete the
standardization of the data, a smoothed energy distribution from the
most-observed secondary standard star, Hya 56 (HD 27962), was used as the
fundamental standard for all cluster stars. Lastly, each wavelength's flux was
normalized with respect to its flux value at 5556\AA.  We will therefore often
refer to our fluxes as spectrophotometric ``colors.''

For the present paper these data are placed onto the spectrophotometry
standard of the CSS. The data quoted in the CSS use a similar scanner-type
system, and are standardized to the system of Hayes \& Latham (1975).  Of the
eight stars in common, two are from Pleiades, four stars from the Hyades, and
two Ap stars from Coma Berenices.  As the data in the CSS for these stars
only extends to $\rm 7100 \AA$, we assume that our data for longer
wavelengths will require the same zero point correction as for neighboring 
shorter wavelengths.

A systematic offset was indeed found between the 1974 reduction and the data of
the CSS.  This offset was easily quantified as being +0.005 mag for the 13 
passbands $\rm > 4000 \AA$, and +0.037 mag for the three passbands at 
3500--3636$\rm \AA$.  The sense of both offsets is that our 1974 reduction 
produced fluxes too bright in the Hayes--Latham (1975) sense.  Our zero point
offsets are in the same sense as those known between the Hayes (1970) and
Hayes--Latham (1975) systems.  The correction for passbands $\rm > 4000 \AA$
are consistent with that given by Hayes and Latham, while that for passbands
in the UV are larger by about 0.02 mag.  Such an offset likely arises
from differences in the true atmospheric extinction from that assumed.
Once corrected, the mutual scatter between our spectrophotometry and that of 
the CSS is $\pm0.016$ mag (1$\sigma$ per color) averaged over all 15 colors,
as seen in Figure~1a.  We adopt these corrections, so that our data are 
formally on the spectrophotometric system defined by Hayes--Latham.

\subsection{Comparison to Other Spectrophotometry}
\label{sec2.3}

Our comparison with the two other spectrophotometric catalogs, SSA and BVJ,
are done using our CSS-corrected data.  We have three stars in common with 
the SSA: Pra 114, 154 and 276.  As can be seen in Figure~1b, a color-by-color
comparison yields relatively poor agreement, $\pm0.046$ per observation for 
all three stars.  There are systematic offsets between our data sets that 
tend to divide into wavelength regimes above and below $\rm 5000 \AA$.

A comparison with the spectrophotometry of BVJ is complicated by the fact that
they calibrated their data to a different standard system than that of Hayes
\& Latham (cf. discussion in BVJ).  As such, we have compared our data with
those of BVJ as follows and summarized in Table~3:  Column (1) gives the color
bandpass for 15 colors (the 4400 color is not used).  Column (2) is the
difference in adopted zero point, BVJ - Hayes--Latham, calculated from the
data given in Table~2 of BVJ.  Column (3) is the mean raw difference, US --
BVJ, with the error of the mean (i.e., $\sigma/\sqrt{N}$) for this raw
difference given in Column (4).  Column (5) is the difference between column
(3) and column (2), giving the observed minus the predicted zero point offsets
(obs--pred) between our data and that of BVJ.  For the three bandpasses below
the Balmer jump, the offset in obs--pred is +0.0133 $\pm$ 0.006 mag (errors of
the mean quoted); while above the Balmer jump it is -0.009 $\pm$ 0.004 mag. 
Overall the difference is only -0.005 $\pm 0.004$ mag.

The difference, star-by-star, {\it after} the mean raw difference (US--BVJ)
has been removed from the comparison, is shown in Figure~1c. Comparison with
Figures~1a and 1b indicate that the scatter between our data that those of
BVJ is larger than that with the CSS data, but smaller than that of Gunn \&
Stryker. A star-by-star comparison between BVJ and US indicates that most of
the difference is systematic, cluster-by-cluster.  For the three bandpasses
below the Balmer jump, the mean difference, US--BVJ is -0.025 mag for 5 Coma
stars (with one star having a difference of -0.050 mag), +0.024 mag for 4
Praesepe stars (one star with +0.061, another +0.041), but less than 0.01 mag
for 5 stars in Hyades and 2 in Pleiades.  Agreement is generally better for
all clusters for the 12 bandpasses redward of the Balmer jump, with typical
differences of 0.01--0.02 mag, and no difference greater than -0.033 mag.

Two stars have been observed by three data sets:  Hya 95 by US, CSS and BVJ;
and Pra 154 by US, SSA and BVJ.  The three-way intercomparison is shown in
Figure~1d.  The agreement among US, CSS and BVJ for Hya 95 is reassuring,
showing scatter per observation of less than 0.01 mag. However, it is clear
from comparing US vs. BVJ and and BVJ vs. SSA for Pra 154 that none of the
observations are in good agreement with the other, although BVJ seem to agree
a little better with SSA than US with BVJ.

Why we agree better with CSS than BVJ or SSA is not obvious.  Of the 8 stars
in common with CSS, all are in the same clusters as used in comparing to BVJ
(Coma, Hyades, Pleiades).  BVJ observed some of their stars in common with us
on more than one night, others just on one night (as in the present survey).
The ``usual'' suspects --- errors in nightly extinction or nightly zero
points, variable stars, just bad luck with random statistics --- are likely,
but not proven.  Fortunately, we have other means of assessing the accuracy of
our data (\S~\ref{sec3}, \S~\ref{sec4}, \S~\ref{sec5}) that include most of
the stars in our sample. As a result, we defer a final discussion of the
errors of our spectrophotometry until all comparisons are made.

\subsection{The Calibrated Data}
\label{sec2.4}

The final calibrated spectral energy distributions (SEDs; relative to
5556\AA) for all 237 stars are given in Table~4 according to the cluster
identification number and the night(s) of the
observation.  The source of cluster identification number is given in the
Table notes. A colon is placed after values with {\it a priori} known 
estimated errors of $\sim$0.03--0.06 mag; an asterisk indicates errors 
$>$ 0.06 mag (which were usually interpolated from adjacent wavelengths).  
This table will also be made electronically available as described at the 
end of this paper.

\subsection{Cluster Color--Magnitude Diagrams}
\label{sec2.5}

The SIMBAD database was searched to obtain V magnitudes, UBVRI colors, {\it
uvby} colors and spectral types for the stars.  From these data we construct
up-to-date color--magnitude (C--M) diagrams for the stars in these clusters
that are in our survey (Figures~2a--g).  Also plotted in the cluster C--M
diagrams is the fiducial zero age main sequence (ZAMS) from Mihalas \& Binney
(1981), adjusted to match the least-evolved main sequence stars as well as
possible.  Based on this adjustment, some of the distances given for the
clusters in Table~1 have been modified, as noted in that table. The ZAMS is
useful for identifying likely main sequence stars for the comparison with
Kurucz models (\S~\ref{sec4}), and for detecting spectral type
misclassifications (Appendix A). The stars specifically identified in
Figure~2 will be discussed in Appendix A and elsewhere in the paper; stars 
used as secondary flux standards are noted with bold numbering.

We initially assume that luminosities and spectral classes of stars are as given
by the MK types from SIMBAD.  However, as can be seen in Figure~2, not all
stars have MK classifications that correspond well to their position in the
C--M diagram.  We therefore modify our assumptions such that any star which
could be sensibly defined as a main sequence star based on its position in the
cluster C--M diagram was treated as such. Separately, most of the clusters
appear to have a binary star population as evidenced by the stars lying
parallel to, but offset from the ZAMS line.

\section{Comparison of Spectrophotometry to Filter Color Systems}
\label{sec3}

The spectrophotometry colors can be used to synthesize the filter colors of
the Str\"{o}mgren {\em uvby} and Johnson BV systems (cf. Breger 1976; Adelman
1981). For Str\"{o}mgren colors, two adjacent bandpasses are averaged to
achieve a closer approximation to the filter effective wavelength.  In the
case of the Johnson BV filters, a convolution was made between these data and
the B and V sensitivity functions (S($\lambda$)) defined by Matthews \&
Sandage (1963) such that:

\begin{equation} 
(B-V)_{1} = 2.5\times 
\log \frac{\int_{0}^{\infty}S_{V}(\lambda)_{1}F(\lambda)d\lambda}
{\int_{0}^{\infty}S_{B}(\lambda)_{1}F(\lambda)d\lambda} .
\end{equation}

\noindent F($\lambda$) represents the flux taken from the present data and the
subscript of 1 indicates that the sensitivity functions with an airmass of 1
were used to determine the B-V color. Table~6 list the observed B--V and {\it
uvby} data as obtained from SIMBAD, together with the differences (observed
minus predicted) between the colors we predict from our spectrophotometry and
the observed colors.  We will find these color ``residuals'' quite
valuable in assessing the accuracy of our data (below and \S~\ref{sec5}).

The tranformations between our spectrophotometry and the filter colors are
given in Table~5, together with the results from a simple linear fit between
predicted colors and observed colors.  As all of our colors imply a magnitude
difference of the given passband with the flux at 5556$\rm \AA$, the
synthesized indices can be simply defined. Seventeen stars do not have 
{\it uvby} data.  Seven stars were found to have significant residuals 
($> 3 \sigma$) with respect to the {\it uvby} color fits:  Com~65 (v--y; u--b);
Com~162 (b--y); Com~183 (v--y; b--y); IC4~83 (b--y); IC4~89 (u--b; b--y);
Per~285A (b--y) and Ple~2181 (a known variable star; v--y; u--b).  The color 
residuals for these stars are foot-noted in Table~6, and these colors are not 
used to defined the transformation between the spectrophotometry and the 
filter colors.

The residuals, in the sense (observed-predicted) are given in
Figures~3a,b,c,d, with stars from the different clusters denoted by different
symbols.  All the comparisons have a slope near unity, with deviations of $\pm
\sim 0.05$ mag from linearity in the color range -0.35 to -0.50 with B--V and
in the range of -0.25 to -0.50 with {\em v-y}. Such deviations are not
surprising, given the avoidance of strong absorption features by the
spectrophotometry bandpasses, and the known sensitivity of the filter indices
in the wavelength region 4000--4900$\rm \AA$ to changes in helium and hydrogen
absorption in this temperature range. The fact that we see little deviation in
linearity in the {\it u--b} comparison is a testament to the strong line
blocking below the Balmer jump which makes any filter in this wavelength range
sensitive to absorption features.

Regarding the two clusters whose secondary stars are suspected variables,
Praesepe and Coma Berenices, we find that their stars follow the average 
filter--SED {\it b--y} correlation to 0.001 mag and -0.001 mag, respectively, 
with scatter of 0.015 and 0.017 mag ($1-\sigma$).  This is quite comparable to
the zero points and scatter found for the other 5 clusters (a range in zero 
point from -0.0004 to +0.0027 and scatter from 0.010 to 0.018 mag).  
Similarly, we find Praesepe and Coma Berenices stars to follow the
filter--SED {\it u--b} zero point to 0.004 to 0.01 mag, with scatter of 
0.026 and 0.021 mag, respectively.  Again, this is comparable to what we
find for the stars in the other 5 clusters (zero points from 0.001 to -0.010
mag; scatter from 0.017 to 0.031 mag).  Thus, we can find little evidence
that the secondary standard stars varied sufficiently during the period of
observation to differentially affect the SEDs of the stars in these clusters
at a level we can detect.

Given the obvious non-linearity in the relation with {\it v--y} and B--V, our
preliminary estimate of the accuracy of our photometry from the filter
comparison comes from the fits to {\it u--b} and {\it b--y}, which sample a
comparable range in wavelength as the other two colors.  As shown in Table~5,
we get a ($1-\sigma$) scatter of 0.023 mag for {\it u--b} and 0.013 mag for
{\it b--y}.  The implied scatter for our data (0.01--0.02) mag compares
favorably with the estimate we get from comparing to the CSS, but is clearly
smaller than our comparisons to BVJ and the SSA.  Lines of $\pm 0.015$ mag are
drawn in Figure~3 to guide the eye relative to the scatter seen.

\section{Comparison to Kurucz Models}
\label{sec4}

\subsection{The Kurucz Models}
\label{sec4.1}

The stellar synthesis models of Kurucz (e.g., 1979, 1992, 1993) have been
widely used over the years, especially as Kurucz has strived to improve their
relationship to real stellar spectra.  For example, Adelman (1981) used Kurucz
(1979) models to estimate temperatures and gravities for certain stars in the
CSS.  We believe that by comparing the models of Kurucz (1993) to our data, we
are making one of the more comprehensive comparisons of spectrophotometry
between models and stars with a large range in temperature. As will be seen, we
also benefit from this comparison, as the models give us an estimate of the
accuracy of our spectrophotometry that is both more extensive and independent 
of the filter colors, and more comprehensive than intercomparisons with other
observers (\S~\ref{sec2}, \S~\ref{sec3}).

We opt to explore only the variation of temperature with respect to the Kurucz
models for the stars in our sample for two reasons.  First, the differences in
[Fe/H] from solar are at the 0.1 dex level for these clusters.  Second, likely
gravity differences among upper main sequence stars of ages 30--1000 Myr are
generally less than 0.5 dex, at the resolution of the grid of models provided
by Kurucz.  Rather than interpolate among the Kurucz models for relatively
small gravity and metallicity differences, we choose instead to explicitly
test for evidence of such differences.  We compare our data to Kurucz models
with [Fe/H] = 0.00, $\rm \log g = 4.0$ and a microturbulent velocity of 2 km
sec$^{-1}$. A spectral resolution of 32$\rm \AA$ resolution was assumed for
this comparison, as most of the data were taken with this resolution.  We take
into account the reddening of each cluster by appropriately reddening the
Kurucz models for the stars in each cluster, using the E(B--V) values given in
Table~1 and a standard Seaton (1979) reddening curve.

\subsection{Ordering of Stellar SEDs by Cluster and Temperature}
\label{sec4.2}

We compare the Kurucz models to all stars in our data which we classify as main
sequence stars either based on their spectral classification or by their
placement in the cluster C--M diagrams (\S~\ref{sec2.5}). The best--fit Kurucz
model to the data is chosen by comparing the full range of Kurucz models with
[Fe/H] = 0.0, log g = 4.0 and varying temperature. For each model comparison we
calculate a least squares fit parameter $s$, defined as $s =
\frac{\sqrt{\sum{res^2}}}{13}$, where $res$ is the difference between our flux
and that of Kurucz only for the 13 passbands with $\rm \lambda > 4000 \AA$;
i.e., above the Balmer jump. The minimum of this distribution for the residual
sum is taken to be the appropriate temperature model.  For this comparison we
specifically exclude the 3 bluest spectrophotometric colors for the well-known
reason that gravity and metallicity effects are far larger for colors defined
by combining passbands above and below the Balmer jump, relative to colors
defined where both passbands are above the Balmer jump. Rather, based on the
best fit to the model for the wavelengths above the Balmer jump, two
separate average residuals (obs--model) are calculated for each star; one
for the 13 ``red'' colors and one for the 3 ``blue'' colors.

In this process we do not {\it a priori} require that the mean (obs--model)
residual for each cluster be exactly zero.  Rather, by letting this value
float freely, we provide an extra check on the accuracy of our comparison.
As expected, the average difference per cluster, (obs--model) for the 3 ``blue''
colors shows signficant scatter (see also below).  Despite that scatter,
significant zero point offsets (at $> 5 \sigma$ significance) of -0.013 and 
-0.017 mag are found only for the Hyades and Praesepe clusters.  Since these 
clusters are the oldest in our sample, somewhat metal-enhanced (Table~1), and
with cooler stars whose SEDs are more affected by metallicity, such a 
small difference is not surprising given our use of solar metallicity models.  
Interestingly, zero point differences at $\le 0.01$ mag level are also seen for
the average of 13 ``red' colors for Hyades, Pleiades, Coma Berenices
and IC~4665 stars.  

Figures~4a--g show the SEDs for the main sequence stars in each cluster,
grouped according to their best-fit Kurucz temperature. For the most part, the
stars of similar temperature have SEDs that match well (indicating that the
comparison worked), with the largest differences in the 3500--3636$\rm \AA$
range, as expected. Figure~5 shows the SEDs of the non-main sequence stars
(luminosity classes I, III, and IV) which were not explicitly compared to
Kurucz models, grouped according to SED.

\subsection{Gravity and Other Effects in the Kurucz Comparison}
\label{sec4.3}

Figures~6a,b graph the absolute V magnitudes of the stars in our sample (using
our predicted distances and reddenings) versus the residuals
(observed--predicted) for the fit of Kurucz models to our data, with the known
(small) zero point differences removed.  Figure~6a plots $M_V$ versus the blue
color residuals; Figure~6b plots $M_V$ versus the red color residuals. 
Note the difference in plotting scale used for the residuals in Figure~6a
versus that used in Figure~6b.

As would be expected, the scatter of the Kurucz models versus the 13 red
colors of our spectrophotometry show no residual dependence on absolute
magnitudes, with an average scatter ($1-\sigma$) of 0.01 mag per star
(Figure~6b).  As equally expected, the scatter of the Kurucz models versus
the 3 UV colors shows highly systematic differences as a function of 
stellar absolute magnitudes.  The most obvious effect is that gravity 
differences among the upper main sequence and subgiant stars strongly affect
the UV residuals.  This is to be expected.  

More interesting, perhaps, are the more subtle differences among the red color
residuals for the lower main sequence stars.  In the range $3.0 < M_V < 2.5$
are seven Hyades stars that progressively deviate from the Kurucz models with
brighter magnitudes (\#'s 38,45,53,68,80,94,103).  Of these stars, two are
known Am stars, one is classfied as a subgiant (!), and the others are either
F0V or F4V.  It is possible that these small, but systematic differences are
caused by increased line absorption symptomatic of the Am phenomenon (might
this also be the source of the apparently mis-applied subgiant
classification?), rather than variability, as no similar variation is seen
for the red colors of these stars. Separately, stars fainter than $M_V = 4.5$
in both Hyades and Coma show a somewhat wider spread in red residuals than do
brighter main sequence stars in these two clusters.  Why this latter effect is
so is not clear, as a gravity effect along should produce a net shift in the
residuals, rather than a wider range.  Perhaps some of the stars are low
mass-ratio binaries, which could effect the model comparsion only in the
reddest colors?

\section{A Quantitative Assessment of Spectrophotometry Errors for Individual 
Stars}
\label{sec5}

The quantitative comparisons among our spectrophotometry, the CSS
spectrophometry, the {\it uvby} and B--V colors, and the Kurucz models all
indicate an accuracy for our spectrophotometric colors of 0.01 - 0.02 mag,
$1-\sigma$. Yet, in quoting $1-\sigma$ errors, we realize that in a data set
as large as ours, $2-\sigma$, $3-\sigma$ and even greater errors must exist if
the distribution of errors is Gaussian.  As these data can be useful in
accurately defining spectral energy distributions of bright stars for use in
stellar population models and for calibrating CCD images, it is highly
desirable that the photometric accuracy for the SED of each star be known.

Fortunately, most of the stars in our sample have both {\it ubvy} colors and
residuals from the Kurucz models.  Suppose that the Kurucz models define
accurate spectral energy distributions for stars in the wavelength region we
have sampled {\it and} the errors in the {\it uvby} filter colors are small
and uncorrelated with our spectrophotometric errors.  If both statements are
true, if we plot the (observed-filter) residuals to the {\it uvby} colors
($\delta$(color)) versus the (observed-model) residuals to the Kurucz models
($\delta$(Model)), these residuals should be strongly correlated.

Figure~7a shows the correlation of $\delta(u-b)$ with $\delta$(Blue Model);
Figure~7b the correlation of $\delta(b-y)$ with $\delta$(Red Model). The term
Red Model applies to the mean residuals for the 13 red colors; similarly, the
term Blue Model applies to the mean residuals for the 3 UV colors.  In the
case of the $\delta(b-y)$ vs. $\delta$(Red Model) a clear correlation exists. 
For 198 stars with (b--y) colors, Kurucz model fits and not otherwise large
color residuals, the correlation coefficient between $\delta(b-y)$ and
$\delta$(Red Model) is 0.55 with a $1-\sigma$ scatter of 0.011 mag. For the
132 stars with $M_V \ge 1.00$ and with good predicted {\it (u--b)} colors, the
correlation coefficient between $\delta(u-b)$ and $\delta$(Blue Model) is
0.298 with a $1-\sigma$ scatter of 0.023 mag.

The fact that we see significant correlations between the color residuals and
the model residuals means that the above hypothesis is reasonably correct: In
comparing to the filter colors and the Kurucz models, we see mainly the errors
in the spectrophotometric colors, while the errors in the filter colors and the
models are much smaller than these.  As a result, we have chosen to tabulate
the values of $\delta(u-b)$, $\delta(v-y)$, $\delta(b-y)$, and $\delta(B-V)$,
$\delta$(Blue Model) and $\delta$(Red Model) for those stars with requisite
data in Table~6, rather than the predicted filter colors.  Inspection of the
values of $\delta(b-y)$ and $\delta$(Red Model) in particular are useful in
choosing stars from our sample with the most accurate spectrophotometry. The
other color residuals and $\delta$(Blue Model) are also somewhat useful in
this regard as a cross-check.

Given this conclusion, it is reasonable to interpret the scatter in the 
correlation of the residuals in Figure~7 as due primarily to the errors in
the spectrophotometric colors.  This means that one can use the size of 
the color and Kurucz residuals to reliably estimate the accuracy of the
SED of each star, taking $1-\sigma$ errors of 0.01 mag for the red colors, and 
0.023 mag for the blue colors as the starting point.  Separately, we find these
error estimates consistent with our estimated errors from our 
spectrophotometric comparison with CSS, but significantly smaller than our
comparisons to SSA (0.046 mag) and BVJ (up to 0.03 mag) (\S~\ref{sec2.3}).

\section{Summary}
\label{sec6}

We present spectrophotometric data in 16 bandpasses from $\rm 3500 - 7800 \AA$ 
for 237 stars in seven nearby open clusters which range in age from 30 to
900 Myr: Hyades, Pleiades, Coma Berenices, $\alpha$ Persei, Praesepe, IC~4665
and M39.  These data are presented in the form of spectrophotomeric colors
(color $\lambda$ = $\lambda$5556 - $\lambda$).  Absolute calibration of the
colors is done by comparing our data with those of Adelman et al. (1989) for
eight stars in common.  External comparisons are then done with repsect
to other published spectrophotometry, {\it uvby} colors, B--V colors and
the predicted spectral energy distributions of Kurucz (1993) models.

Both the filter color comparison and the comparsion to Kurucz models yield
accuracy estimates for our spectrophotometry consistent with what we find
in our comparison with Adelman et al.:  a $1-\sigma$ accuracy of
0.01 mag for the 13 colors formed with passbands redward of the Balmer jump, 
and 0.023 mag for the 3 UV colors.  In contrast, comparison to the published
spectrophotometry of Gunn \& Stryker (1983) and Bohm-Vitense \& Johnson (1977)
indicates larger errors (0.03 - 0.045 mag).

The comparison to Kurucz models was made for main sequence stars with values
of log gravity and [Fe/H] fixed at 4.0 and solar, repspectively.  Gravity
differences and absorption line differences among the stars, more finely 
than can be measured with the standard Kurucz grid, are seen when residuals
(obs minus model) are separately tabulated for the 3 UV colors and the 13
red colors.  While the residuals for the red colors show no net gravity
effect, the UV colors show a strong gravity effect, as would be anticipated
from colors that measure the strength of the Balmer jump in B-F stars.

We show that it is reasonable to assume that, compared to the spectophotometric
errors, the errors in {\it (u--b)}, {\it (b--y)} and the Kurucz models are
small, so that the residuals $\delta(u-b)$ are correlated with the Kurucz UV
residuals ($\delta$(Blue Model)), and the residuals $\delta(b-y)$ are
correlated with $\delta$(Blue Model).  These residuals can be used to identify
those stars for which the spectrophometric colors are the most reliable, and
those which are less reliable.  Thus, while we quote $1-\sigma$ accuracies for
our data, we can also determine which stars have spectrophotometry within
$1-\sigma$ accuracy, $2-\sigma$, etc.  This list should therefore be valuable
to those who would like to have accurate spectrophotometric color standards
for calibration of imaging and spectroscopic data.

Table~4 and Table~6 will be made electronically accessible via both
anonymous ftp (\@samuri.la.asu.edu = IP \#129.219.144.156) and through
the Astronomical Data Center.

\acknowledgements

These data would not exist without the hard work and perserverence of Lee
McDonald, for which we sincerely thank him.  LC was supported in part by a
NASA Space Grant assistantship; DB was supported in part by NASA Grant
GR~6309.194A.  We thank Linda Stryker and Per Aannestad for helpful 
suggestions.  We also thank the anonymous referee for perceptive comments 
on the first version of this paper that were particularly helpful. The SIMBAD 
database was used to obtain much useful data for these stars.

\appendix

\section{Discussion of the Peculiarities Individual Stars}

A number of stars are noteworthy in terms of how their MK classifications
and SEDs correspond to their positions in the C--M diagrams (Figure~2).  
We discuss these stars on a cluster-by-cluster basis.

{\bf Hyades:} Hya~126 (= HD~31236) falls below the main sequence despite
being classified F3 IV.  While its predicted temperature,
$\sim 7500$ K, is consistent with an F classification, its SED gives one of
the worst comparisons to the Kurucz models.  Hya~112 (= HD~30210) has one of
the best comparisons to Kurucz models, but also appears a bit
underluminous for its color.  As this star is classified Am, possible
variation in color and magnitude could produce such an effect.  Hya~72 
(= HD~28319) lies well above the main sequence and is classified as A7III.  
Yet, its SED is similar to those of main sequence stars of similar temperature,
such that this star was included in the Kurucz comparisons.  Hya~72 could 
be a blue straggler, given its position relative to the main sequence
turn-off.

{\bf Pleiades:} Ple~563 (= HD~23338), classified B6IV, has an SED that is
in very good agreement with the other subgiants in this cluster at wavelengths
above 4000$\rm \AA$, but is signficantly more luminous below the Balmer jump.
Ple~2181 (= HD~23862) is a known variable which affects its filter color 
comparisons. Ple~1375 (= HD~23629, A0V) has a similar estimated
temperature as Ple~817 (= HD~23432, B8V), despite being both redder and
less luminous in the C--M diagram (cf. Figure~2b). These two stars also differ
significantly in their fluxes below the Balmer jump, with Ple~1375 being
0.3 mag redder than Ple~817. As the SED of Ple~817 is more in accord
with Kurucz models, this would indicate Ple~1375 is a likely binary,
consistent with its offset position from the main sequence.

{\bf Praesepe:} Pra~265 (= HD~73666) is signficantly bluer than, and
offset from, the well-defined main sequence of this nearly 1 Gyr old cluster.
With an F6V classification it could either be an interloper in the
cluster, or a blue straggler.  Separately, five of the stars in our sample are
nominally classified as giant stars, but clearly lie within the evolved main
sequence of this cluster (224 = HD~73618; 276 = HD~73711; 279 =~HD 73709; 292
= HD~73729; and 328 = HD~73785).  These must either be misclassifications or
non-member interlopers. Pra~142 (no HD number) and Pra~275 (no HD number) are
among the faintest and coolest main sequence stars in our sample.  They are
also likely binary stars, as they lie well above the fiducial single star main
sequence of this cluster.  This is consistent with the SED comparison to 
Kurucz models, which implies the SEDs of these two stars are different from
those of a single star.

{$\alpha$ \bf Persei:} Most of stars in this cluster are classifed as Ap.  
Per~868 (= HD~21479) is classified as A1IV yet it lies near the main sequence 
(Figure~2d).  Of the five most luminous stars extending above the main 
sequence (not counting Per~605, a supergiant), only one is formally classified
as a peculiar giant (Per~985 = HD~21699, B8IIImnp).  All of the other stars 
at the tip of the evolved main sequence are classified as main sequence, 
B3--B6, with one being B6Vn.

{\bf Coma Berenices:}  One third of the stars in this cluster are classified
as A--peculiar (p, n or m), the largest fraction of any of the seven clusters.
Two of the stars in common with CSS (Com~146 = HD~108662 and Com~160 =
HD~108945) are noted by Adelman (1981) as being variable.  Others are noted as
variable in the SIMBAD data base.  The two obvious evolved stars in this
cluster (Com~125 = HD~108283 and Com~91 = HD~107700) do not have luminosity
classifications.  Interestingly, the SED of Com~91 (F2) matches well with the
SED of the supergiant star Per~605 (F5Iab), despite being in very different
positions relative to the main sequences in their respective clusters.

{\bf IC~4665:} This cluster has the most poorly-defined main sequence of any
of these seven clusters.  It is not {\it a priori} clear whether the scatter
along the main sequence is due to a strong binary star population, poor
data or field star contamination.  One star, IC4~83 (no HD number) is defined 
as a double system by SIMBAD, but is not otherwise much offset in the C-M 
diagram from the nominal single star main sequence.

{\bf M39:}  Only nine stars are observed in M39, the fewest of any
cluster in our sample.  Three of the 6 main sequence stars are classified Ap.

\clearpage

\noindent {\bf Figure Captions}

\vspace*{3mm}

\noindent {\bf Figure 1.}  (a) A comparsion of the SEDs of eight stars
to the data in the Catalogue of Stellar Spectrophotometry (Adelman et al.).
The stars used are indentified in the legend, and the residuals are in 
magnitudes.  Dotted lines are draw at $\pm$0.015 residuals for guidance.
(b) A comparison of the SEDs of three stars to the data of the
Stellar Spectrophotometry Atlas (Gunn \& Stryker).  Note the much larger
scatter, both random and systematic, in (b) than in (a).  (c) A comparison
of the SEDs of 16 stars in common with the data of Bohm-Vitense \& Johnson.
Here the scatter is intermediate between (a) and (b).  (d) A three-sided
comparison of spectrophotometry for two stars observed by three separate 
observers.  Stars and observers noted in the figure.

\noindent {\bf Figure 2.} V vs. (B--V) color-magnitude diagrams for the
stars observed in each cluster.  Stars are distinguished by MK spectral
type (see figure legend) where available or by estimated luminosity classes 
based on their positions in these diagrams.  Specific stars of interest that
are discussed in the text are indicated using the cluster numbers given
in Table~5.  The Zero-Age-Main-Sequence from Mihalas \& Binney (1981) is also 
given in each C-M diagram for guidance.  (a) Hyades; (b) Pleiades; (c) Coma
Berenices; (d) $\alpha$ Persei; (e) Praesepe; (f) IC~4665 and (g) M39.

\noindent {\bf Figure 3.} The residuals of linear fits of 
({\it u--b}) (a), ({\it b--y}) (b), ({\it v--y}) (c) and (B--V) (d) colors 
with equivalent estimates from the SED data.  The fits used are given
in Table~4.  Data for stars from different clusters are plotted with 
different symbols, as given in the legend. Residuals are plotted in the 
sense of (observed minus SED).

\noindent {\bf Figure 4.} The SEDs of the main sequence open cluster stars, 
arranged by cluster and within clusters, matched by similar temperatures as 
indicated by the comparison with Kurucz models.  In contrast to the other
figures, the SEDs are plotted here as flux ratios, rather than as magnitudes.
Stars in each group are identified, as well as estimated temperature range
of the stellar grouping.  

\noindent {\bf (a)} Hyades: 8750K (56), 
8000K--8250K (47,72,82,95,112,54,104,108),\\
7500K--7750K (24,30,55,60,67,83,84,111,126,141,74,123),\\
7000K--7250K (11,20,32,53,100,38,45,68,80,89,103),
6750K (34,35,36,44,51,78,85,94,101,128),\\
6500K (57,77,81,86,121,124), 
6250K (29,31,40,48,52,59,62,65,66,75,88,105,118,119,146),\\ 
5750K--6000K (23,39,50,73,97,102,106,58).

\noindent {\bf (b)} Pleiades: 14000K--15000K (1823,541), 
12000K--13000K (859,2263,2425),\\
11000K--11500K (1375,817), 9750K--10500K (801,1234,1380),\\
9000K--9250K (1028,2220,2866,1397,1431),
8500K--8750K (717,1425,2289,153,1876),\\
8250K (232,804,1384,2415), 7250K--8000K (157,1084,344,1407,2195).

\noindent {\bf (c)} Praesepe: 9750K (265), 
8000K--8250K (50,114,207,224,279,328,284,300,348,375,445,276),\\
7750K (45,143,203,229,204), 7500K (150,154,284,292,340), 
6750K--7000K (227,239,146),\\ 
6250K--6500K (142,275,268).

\noindent {\bf (d)} $\alpha$ Persei: 16000K--20000K (557,675,774,835), 
14000K--15000K (383,810,955,904),\\
12000K--12500K (831,735), 11000K--11500K (581,1082,692),\\
9500K--10000K (423,868,612,775),
9250K (167,575,639,756,817,875),\\
8500K--9000K (651,780,386,625), 8000K (285A).

\noindent {\bf (e)} Coma Berenices: 10000K (146), 
8500K--9000K (10,130,107,160),\\
8000K (60,62,68,139,144),
7500K--7750K (82,109,104,145), 7000K(19,36,49),\\
6750K (86,90,101,118,183),
6500K (53,58,65,114,162), 6250K (76,85,97,111).

\noindent {\bf (f)} IC~4665: 19000K (58), 
16000K--17000K (49,73,62,105), 13000K--15000K (82,23,32),\\ 
10000K--11500K (81,102,22), 9500K-9750K (43,76), 
8500K-8750K (27,50,83,67,89),\\
8000K--8250K (63,118,39,66), 7500K--7750K (65,51).

\noindent {\bf (g)} M39: 11000K (26), 10500K (1,23), 
10000K (33,40A), 9750K (24,38), 9500K (5,17).

\noindent {\bf Figure 5.}  The SEDs of the non-main-sequence stars in our
data set, plotted as a function of approximate ascending temperature.
Note the similarity of the SEDs of Com~91 and Per~605, despite their 
difference in MK classification (Appendix A).

\noindent {\bf Figure 6.} The relation of absolute V magnitude with
the residuals from the fit of Kurucz log = 4, [Fe/H] = 0.00 model
to the SEDs in which only temperature is varied. (a) The average of the 3
UV colors, termed ``blue model residuals;'' (b) The average of the 13 
red colors, termed ``red model residuals.''  Note that the y-axis scale is
6.5 smaller for (b) than for (a).  Symbols for different cluster stars as
in Figure 3.

\noindent {\bf Figure 7.} The residuals of predicted {\it uvby} colors 
plotted versus the model residuals from the Kurucz model fits.  (a)
$\delta(u-b)$ versus the ``blue model residuals.'' (b) $\delta(b-y)$
versus the ``red model residuals.''  Lines drawn are the average fit obtained
by interchanging X and Y parameters and taking the average of the fits.
Symbols for different cluster stars as in Figure 3.

\clearpage

\begin{table}
\begin{center}
\caption{Table of Selected Clusters}
\vskip 2ex
\begin{tabular}{llrrrccc}
\hline
\multicolumn{1}{|c|}{Cluster} & \multicolumn{1}{c|}{Other} & 
\multicolumn{1}{c|}{l\tablenotemark{a}} 
& \multicolumn{1}{|c|}{b\tablenotemark{a}} 
& \multicolumn{1}{c|}{d\tablenotemark{b}} 
& \multicolumn{1}{c|}{log age\tablenotemark{b}}
& \multicolumn{1}{c|}{E(B-V)\tablenotemark{b}} 
& \multicolumn{1}{c|}{[Fe/H]\tablenotemark{b}} \\ 
\multicolumn{1}{|c|}{Name} & \multicolumn{1}{c|}{Name} 
& \multicolumn{1}{c|}{($\deg$)} & 
\multicolumn{1}{c|}{($\deg$)} & \multicolumn{1}{c|}{(pc)} 
& \multicolumn{1}{c|}{(yr)} & 
\multicolumn{1}{c|}{} & \multicolumn{1}{c|}{} \\ \hline
\hline
Praesepe&M44&205.5&+32.5&160/180\tablenotemark{c}&8.92&0.03&+0.07\\
Hyades&\nodata&180.1&-22.4&43&8.85&0.00&+0.12\\
Coma Berenices&Mel 111&221.1&+84.1&80/90\tablenotemark{c}&8.66&0.00&-0.03\\
M39&NGC 7092&92.5&-2.3&290&8.30&0.06&\nodata \\
Pleiades&M45&166.6&-23.5&130/180\tablenotemark{c}&8.11&0.06&+0.12\\
IC~4665&\nodata&30.6&+17.1&340/470\tablenotemark{c}&7.89&0.17&\nodata \\
$\alpha$ Persei&Mel 20&147.0&-7.1&160/230\tablenotemark{c}&7.61&0.10&+0.10\\
\hline
\tablenotetext{a}{Data taken from Alter, et al (1958)}\\
\tablenotetext{b}{Data taken from 5th Ed. of Lund-Strasbourg Catalogue (1987)}\\
\tablenotetext{c}{These distances are adjusted from the placement of the 
ZAMS line (\S 2)}\\
\end{tabular}
\end{center}
\end{table}

\clearpage

\begin{table}
\begin{center}
\caption{Table of Observations}
\vskip 2ex
\begin{tabular}{llll}
\hline
\multicolumn{1}{|c|}{No.} & \multicolumn{1}{c|}{Date} & \multicolumn{1}{c|}
{Clusters observed} & \multicolumn{1}{c|}{Standard Star(s)} \\ \hline
\hline
1 & July 17/18, 1973 & M39 & M39~33\\
2 & Aug. 14/15	& Pleiades&Ple~785\\
3 & Aug. 15/16 & M39,$\alpha$ Persei&Per~605\\
4 & Sept. 15/16	& Pleiades&Ple~785\\
5 & Sept. 16/17	& Hyades&Hya~56\\
6 & Sept. 17/18	&$\alpha$ Persei,Pleiades &Ple~785\\
7 & Sept. 18/19	& Pleiades&Ple~785\\
8 & Sept. 20/21	& Hyades&Hya~56\\
9 & Oct. 12/13	& Hyades&Hya~56\\
10& Oct. 13/14	& Hyades&Hya~56\\
11& Mar. 8/9, 1974 & Hyades,Praesepe,Coma B.&Hya~56\\
12& Mar. 9/10	& Coma B.&Com~60\\
13& Apr. 10/11	& Praesepe&Pra~300\\
14& May 6/7	& Coma B.&Com~60\\
15& May 7/8	& Praesepe,Coma B.,IC~4665 &Pra~300,Com~60\\
16& Aug. 9/10	& IC~4665&IC4~62\\
17& Dec. 29/30	& Hyades,Pleiades,$\alpha$ Persei &Hya~56\\
\hline
\end{tabular}
\end{center}
\end{table}

\clearpage

\begin{table}
\begin{center}
\caption{Comparison to Bohm-Vitense--Johnson Calibration}
\vskip 2ex
\begin{tabular}{ccccc}
\hline
\multicolumn{1}{|c|}{Passband} & \multicolumn{1}{c|}{BVJ Calib} & 
\multicolumn{1}{c|}{Mean(US--BVJ) (obs)} & \multicolumn{1}{c|}{Mean Error} &
\multicolumn{1}{c|}{Obs--Pred} \\ \hline \hline
3509 & 0.047 & 0.071 & 0.0100 & 0.024 \\
3571 & 0.040 & 0.053 & 0.0086 & 0.013 \\
3636 & 0.025 & 0.028 & 0.0075 & 0.003 \\
4032 & 0.016 &-0.005 & 0.0067 &-0.021 \\
4167 & 0.017 &-0.013 & 0.0055 &-0.030 \\
4255 & 0.012 &-0.006 & 0.0006 &-0.018 \\
4566 & 0.002 & 0.011 & 0.0067 & 0.009 \\
4785 & 0.013 & 0.023 & 0.0056 & 0.010 \\
5000 & 0.012 &-0.014 & 0.0072 &-0.026 \\
5263 & 0.012 &-0.006 & 0.0035 &-0.018 \\
5882 & 0.000 & 0.001 & 0.0034 &-0.001 \\
6370 &-0.003 &-0.004 & 0.0054 &-0.001 \\
6800 &-0.006 &-0.001 & 0.0054 & 0.005 \\
7100 & 0.002 & 0.005 & 0.0033 &-0.003 \\
7850 & 0.003 &-0.014 & 0.0074 &-0.017 \\ \hline
\end{tabular}
\end{center}
\end{table}

\clearpage

\setcounter{page}{33}
\begin{table}
\begin{center}
\tablenum{5}
\caption{Synthetic Passbands Used to Match Filter Color Systems}
\vskip 2ex
\begin{tabular}{lcllll}
\hline
\multicolumn{1}{|c|}{System} & \multicolumn{1}{c|}{Color} & \multicolumn{1}
{c|}{Synthetic Indices} & \multicolumn{1}{c|}{m} 
& \multicolumn{1}{c|}{b} & \multicolumn{1}{c|}{$\sigma$} \\ \hline
\hline
Johnson & B-V & Matthews \& Sandage (1963)& 1.072 & 0.433 & 0.021\\
Str\"{o}mgren &{\em u-b}& 3500 - (4565+4785)/2 & 1.005 & 0.125 & 0.024\\
Str\"{o}mgren &{\em b-y}& (4565+4785)/2 & 0.934 & 0.153 & 0.013\\
Str\"{o}mgren &{\em v-y}& (4036+4167)/2 & 0.891 & 0.415 & 0.032\\
\hline
\end{tabular}
\end{center}
\end{table}

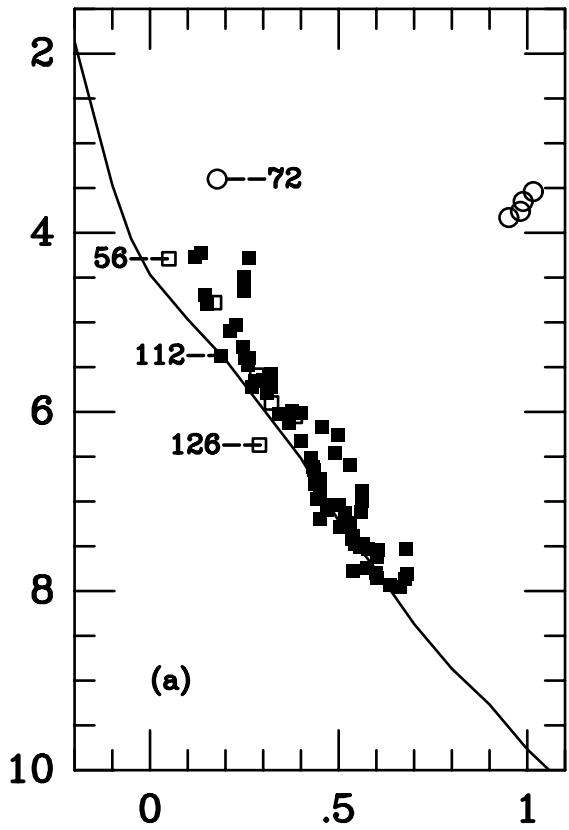
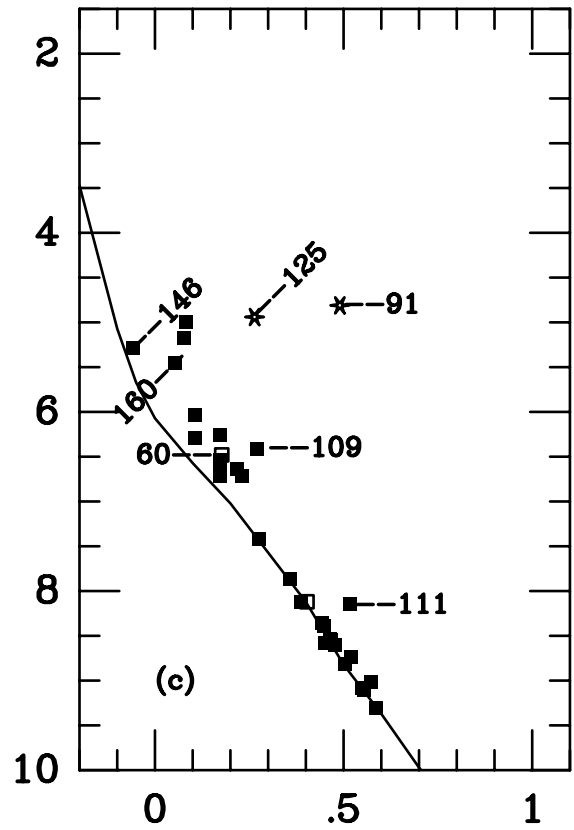
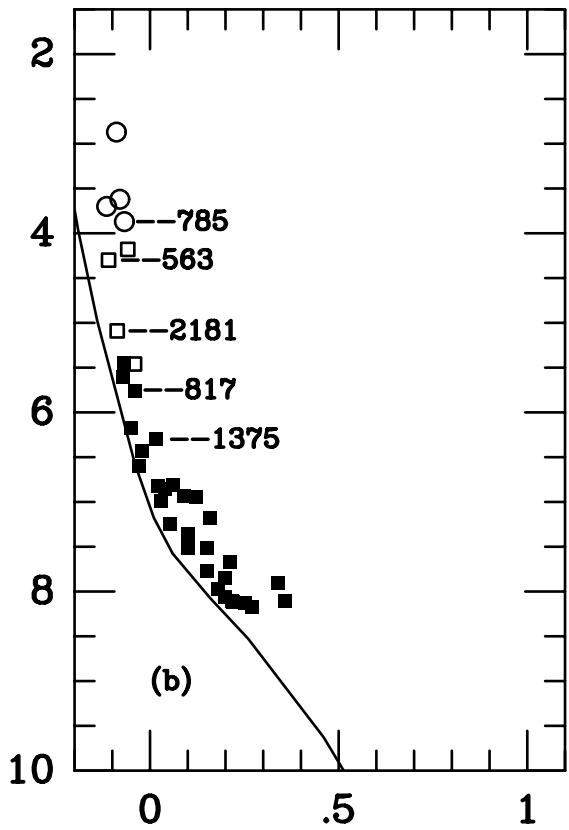
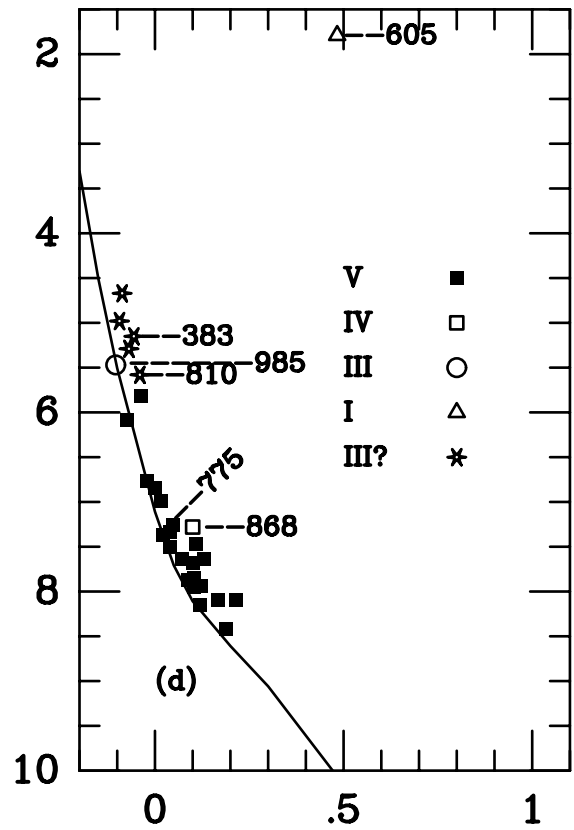
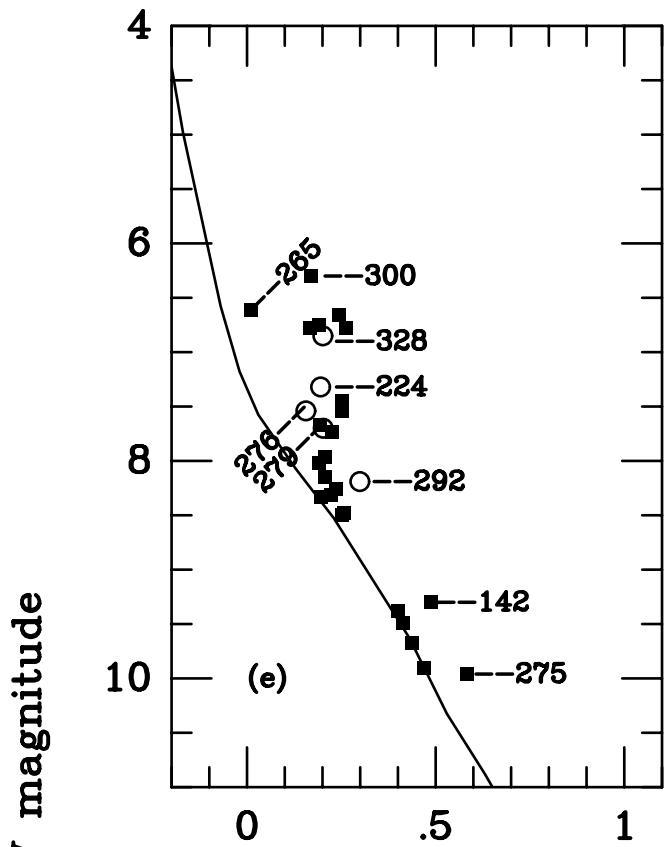
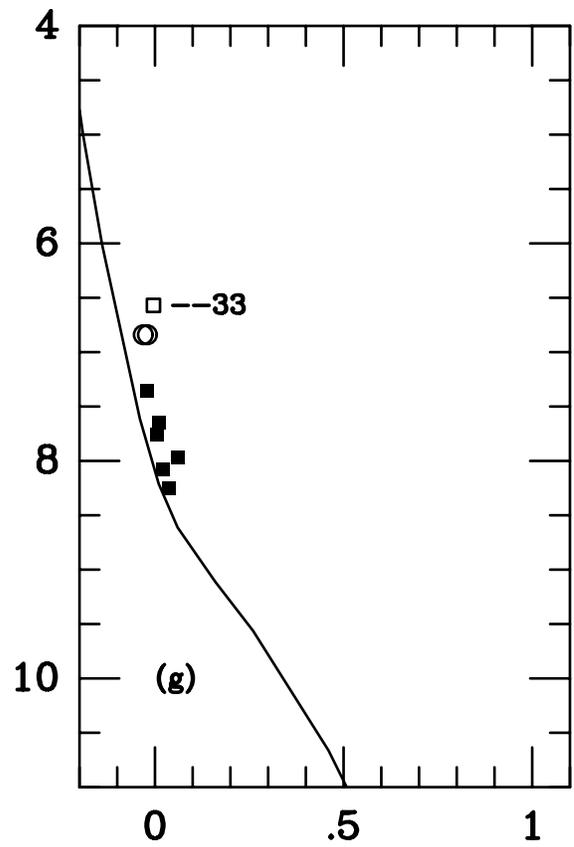
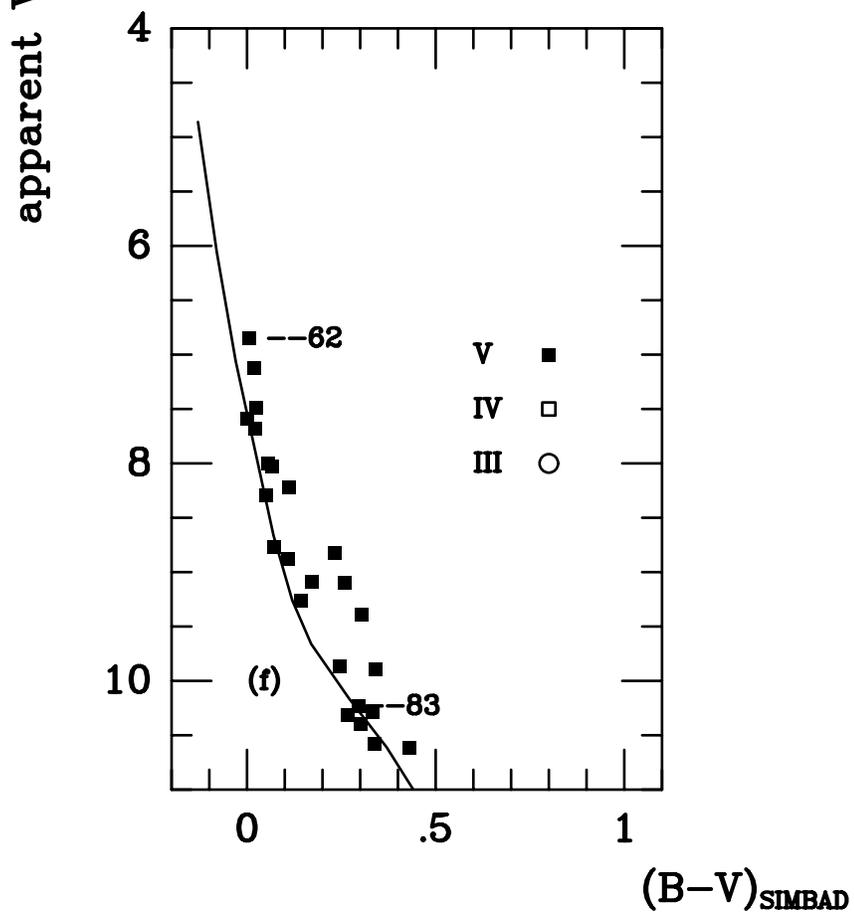
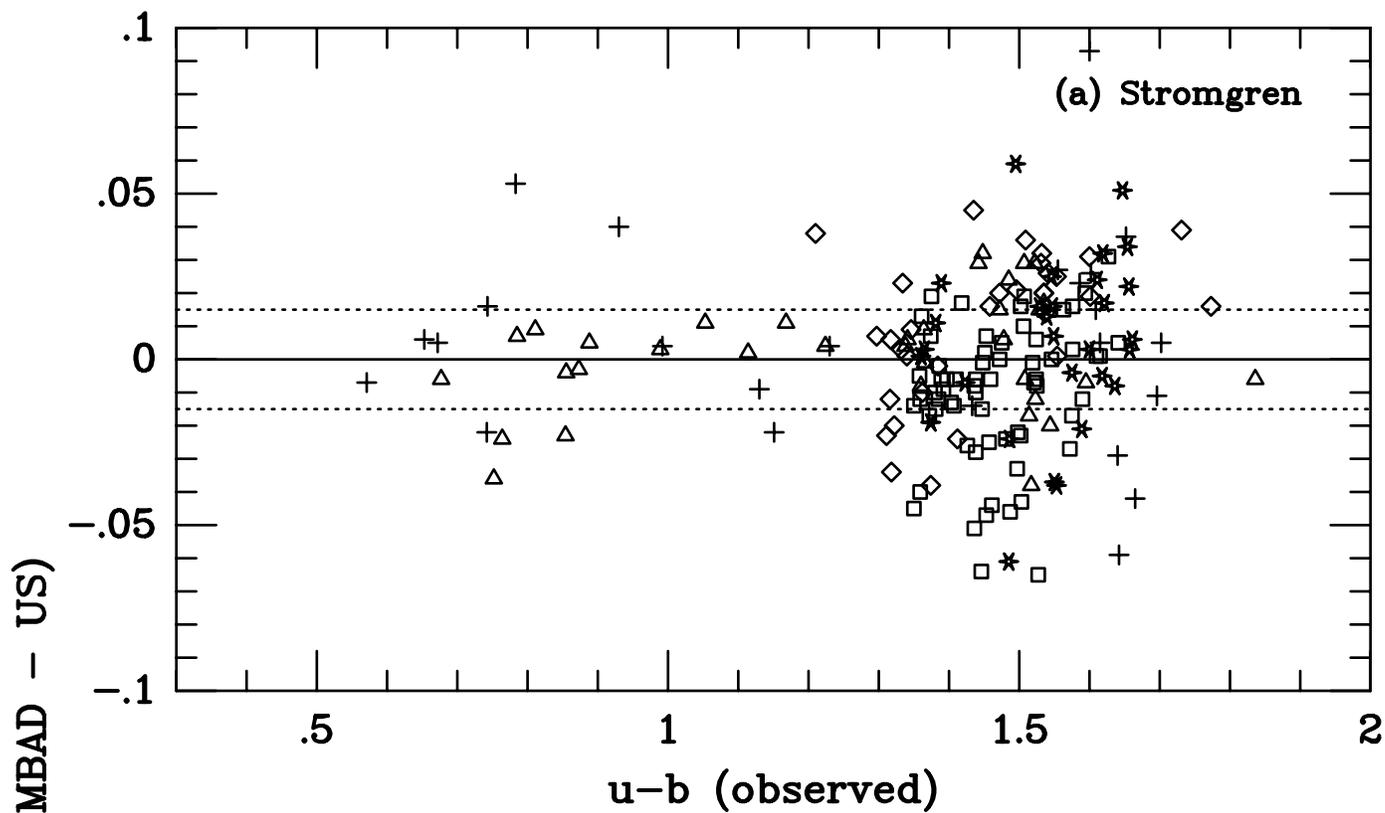
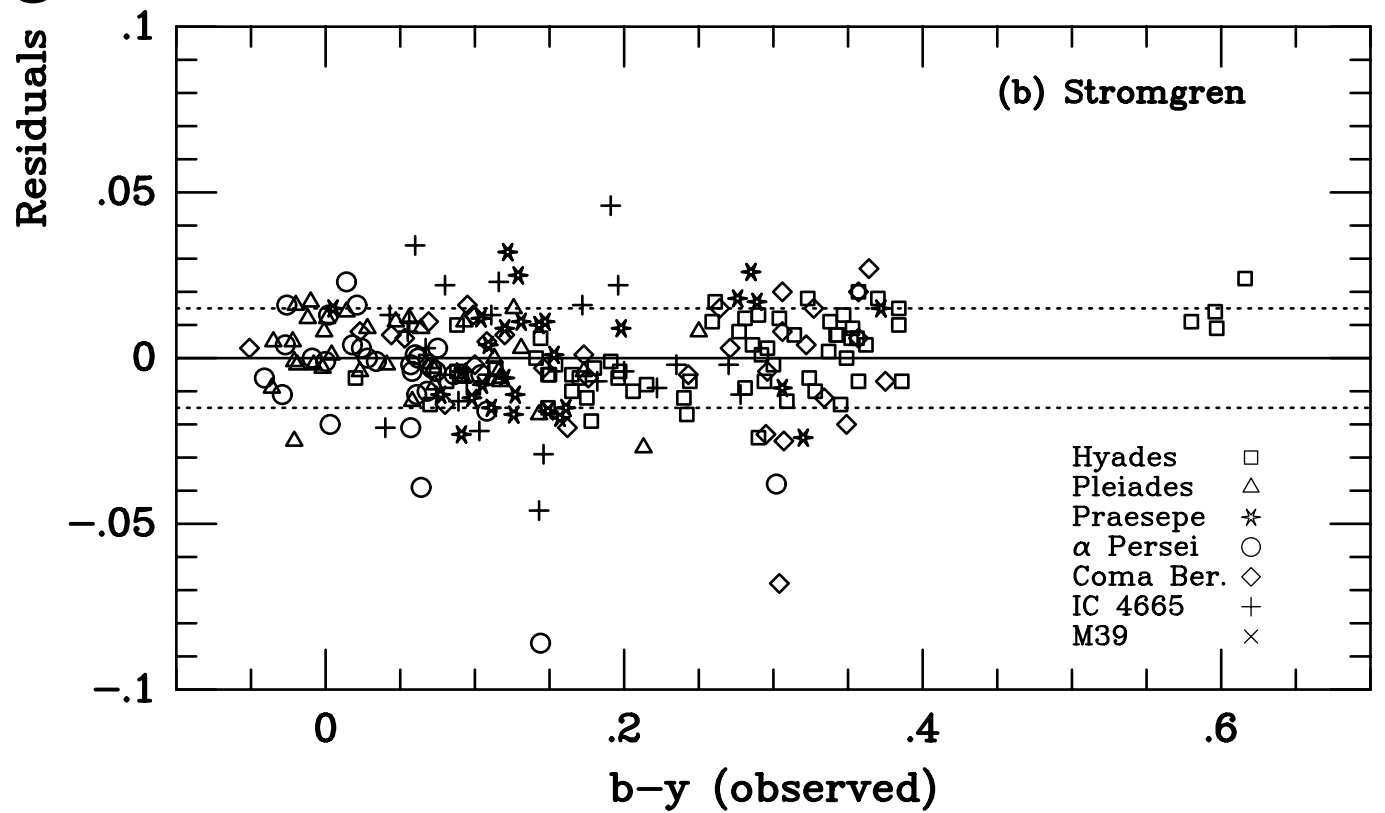
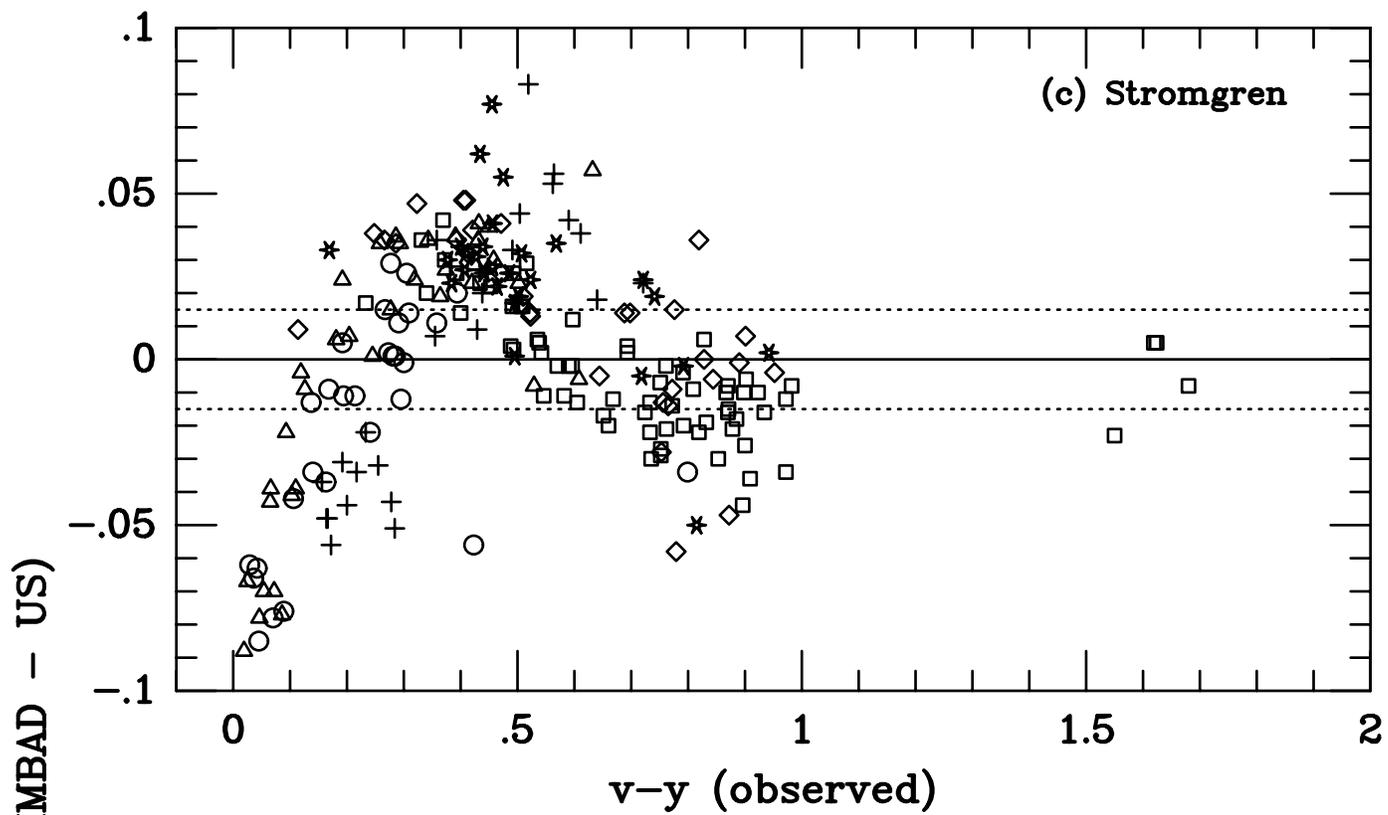
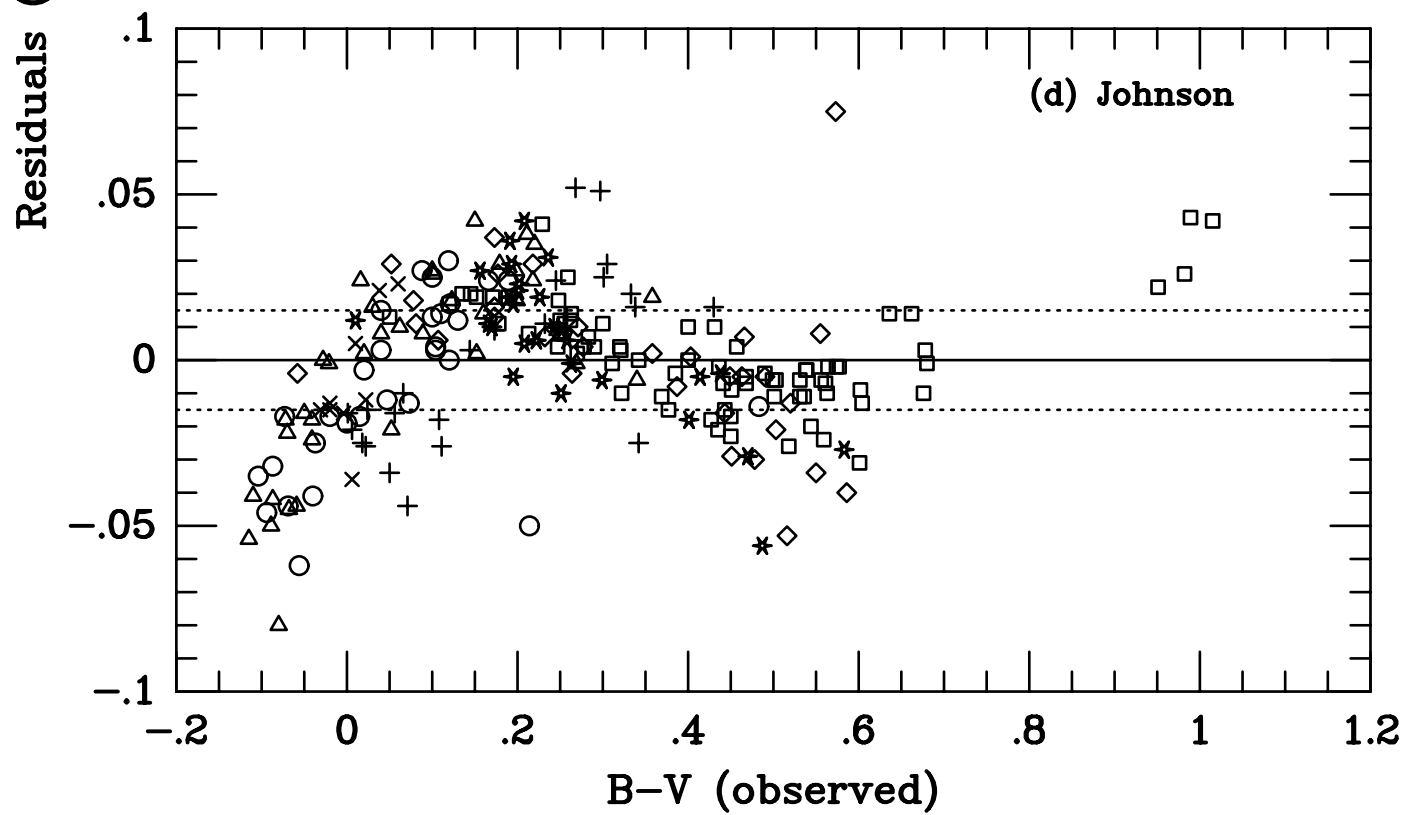
\end{document}